# When Politicians Tweet: A Study on the Members of the German Federal Diet

**Mark Thamm**

GESIS - Leibniz Institute for the
Social Sciences
Unter Sachsenhausen 6-8,
50667 Cologne, Germany
Mark.thamm@gesis.org

**Arnim Bleier**

GESIS - Leibniz Institute for the
Social Sciences
Unter Sachsenhausen 6-8,
50667 Cologne, Germany
Arnim.Bleier@gesis.org

## Abstract

In this preliminary study we compare the characteristics of retweets and replies on more than 350,000 messages collected by following members of the German Federal Diet on Twitter. We find significant differences in the characteristics pointing to distinct types of usages for retweets and replies. Using time series and regression analysis we observe that the likelihood of a politician using replies increases with typical leisure times while retweets occur constant over time. Including formal references increases the probability of a message being retweeted but drops its chance of being replied. This hints to a more professional use for retweets while replies tend to have a personal connotation.

## ACM Classification Keywords

H.3.4 Social networking, H.1.2 Human information processing

## General Terms

Human Factors

**Introduction**

Using social media platforms is a growing trend in political discourse. In 2008 the US presidential campaign has been termed "the Facebook Election"' and a question on "What 140 characters reveal about political sentiment"' [1] first answered for the 2009 German federal election started a run on predicting results with Twitter.

While a simple quantitative relationship between messages generated by the supporters of a political party and actual pooling results has raised doubts [2], more subtle questions regarding the political sphere on social media platforms where brought into the discourse. A common subject of study in this field of research is the microblogging service Twitter – a widely used communication tool equally popular with politicians. Twitter offers its users the opportunity to share short messages (tweets) and to respond to tweets (reply) or simply to forward a tweet (retweet). Based on a network analysis Conover et al. [3] observed that retweet graphs reproduce the known partisan segregation in the offline world while via the reply graph ideologically opposed individuals interact with each other. Another stream of research oriented towards information diffusion builds on regression analysis [4,5] to determine parameters that make tweets more likely to be retweeted.

With two different modes of responding (retweet and reply) on Twitter the question arises if they serve different purposes. If not, we would expect high covariance between the numbers of retweets and replies, conditioned on the total number of tweets. Furthermore, we would expect the same features increasing or decreasing the chance of receiving replies and retweets alike. We here report on two initial observational studies to shed some first light on this question, beginning with the description of the underlying data and its collection. In the first study we look at variations in the communication behavior of the Members of the Parliament[1] themselves over time. We then build on simple regression analysis to gain some insights into the role played by textual features in messages from MPs for the likelihood of being retweeted or replied by the public. We conclude with a discussion of our late findings but also shortcomings of our work in the current stage.

**Data Collection**

The German Federal Diet (Parliament or Bundestag) is one of the two chambers in Germany's bicameral legislature and the central organ of legislative power. The 17th federal election campaign has already been studied via twitter [1] and we choose the elected as objects for our investigation. We started by manually compiling a list containing the usernames of Members of the German Federal Diet on Twitter. With aid for example from dedicated campaign monitoring sides[2] 196 of 620 MPs [6] could be identified having an active Twitter account in November 2011. We now use Twitter's streaming APIs[3] to receive messages directly from these users as well as the retweets of and replies to their messages. In addition we also record messages from other users mentioning one of the MPs on our list. In doing so we already collected a total of 350,000 tweets in in the first year from November 2011 to November 2012.

---

[1] In the remainder referred to as MPs.

[2] E.g.: http://wahl.de

[3] Twitter's streaming APIs imposes a limit on the number of messages to be received. Since we restrict ourselves to the monitoring of 196 accounts we are confident that we remain below that cap.

To approach our research objective we generated two separate datasets from our first years recordings, with the first one covering tweet activities by the MPs themselves and the second covering the responses to the MPs tweets by any arbitrary user. To that end we extracted the retweet and reply-connections. In case of retweet connections we used text similarity between messages as criteria to identify the retweet graph, in case of the reply connection we relied solely on Twitter's API. With this information we generated the first dataset of the MPs' tweeting activities knowing which of them were retweets or replies. The second dataset was generated analogously for reactions to the MPs' messages by the public; thus, including the information which of these have been retweeted or replied to (see Table 1).

| Dataset | total | retweet | reply |
|---|---|---|---|
| all tweets | 350,292 | 74,691 | 114,388 |
| by MPs | 96,787 | 18,091 | 28,441 |
|  |  | **retweeted** | **replied** |
| by MPs | 96,787 | 24,028 | 34,334 |

**Table 1** All retweets and replies within the total corpus. 2. Tweets only by MPs containing retweets and replies. 3. Tweets only by MPs and the number of them that have been retweeted or replied

## Temporal Variation

In this section we look at potential variations in the communication behavior of the MPs over time. Our goal is to see if the proportional relationship between retweets and replies is constant over time or exhibits some characteristic variations. To that end we accumulated the messages in our first dataset – the tweets by the MPs themselves – over the days of the week. In the next step we looked at the distribution of the tweets over the course of the day.

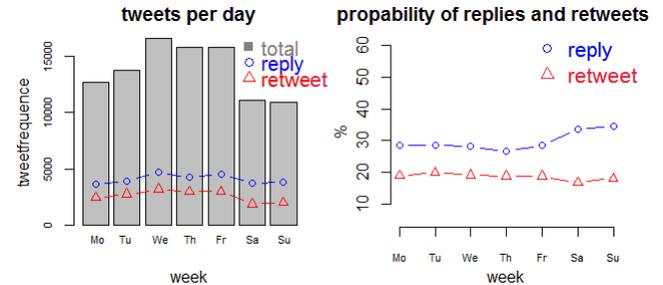

**Figure 1** Left: Aggregated amount of tweets, retweets and replies over the days of the week. Right: The probability of retweets and replies given the total number of tweets by MPs.

In the frequency distribution of tweets by MPs over the days (Figure 2 left) we see that replies are steadily more common than retweets, with the highest tweet activity occurring at the middle of the week. The probability of a reply or retweet given the daily total number of MPs tweets (Figure 2 right) indicates an increasing chance for replies towards the end of the week while the chance for retweets stays far more steady. The distribution of tweets over hours (Figure 3) displays an even sharper disparity in the usage pattern for retweets and replies.

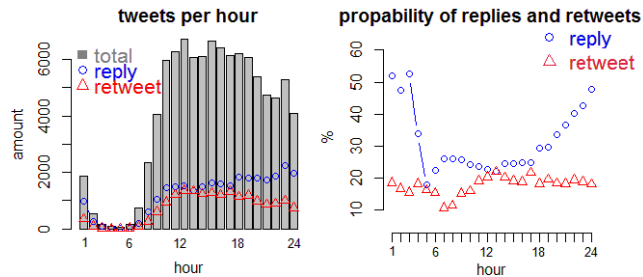

**Figure 2** Left: Histogram of tweets, retweets and replies over the hours. Right: The probability of retweets and replies given the total number of tweets at a certain hour.

The chance of a tweet by a MP being a reply increases in the night while the chance of a MPs' tweet being a retweet follows a steadier path as we have seen in the frequency distribution over the week. This evidence suggests that within the communication behavior of MPs replies and retweets serve different purposes that are time-variant. While the retweet activity is more constant, reply activity tends to increase towards evening and also slightly on weekends. In both cases we see a growth of replies associated with typical leisure times.

**Evidence from the Content**

In this section we look at the content dependence of tweets by MPs to receive attention. It is our goal to see which features make MPs' tweets more likely to be retweeted and which make them more likely to get replies. We employed logistic regression to infer the weight of features and verified them to gain some confidence.

We started by composing a list of features (Table 2) to extract from tweets covering three aspects: emotions, personal orientation and (reference-) link syntax. The features for the emotional aspect follow previous work [5,7] in using a simple dictionary based approach. We employed the Leipzig Affective Norms for German (LANG) [8] to measure valence, arousal and concreteness as well as SentiWS [9] to measure positive and negative strength. Features covering personal orientation capture the mentioning of members of other parties, members inside the group of MPs and the use of pronouns [10] as a measure of personal involvement. The third aspect concerning the link syntax [4,11,12] considers the presence of hashtags, URLs as well as the mentioning of any other user.

| Variable | Description | Range |
|---|---|---|
| negative | SentiWS | 0 .. 5 |
| positive | SentiWS | 0 .. 5 |
| concreteness | LANG | 0 .. 5 |
| Arousal | LANG | 0 .. 5 |
| valence | LANG | 0 .. 5 |
| emoticons | number of used emoticons | 0 .. n |
| pronoun | personal/possessive pronouns | 0 .. n |
| Parties | accounts of competitive parties | 0 .. n |
| Members | referred accounts from MPs | 0 .. n |
| accounts | referred arbitrary accounts | 0 .. n |
| htags | included tags | 0 .. n |
| url | includes URL | 0, 1 |

**Table 2** Regression variables and ranges

We again took all tweets written by the MPs (third row in Table1), but used two logistic regression models to induce a mapping via our features to the chance of receiving a response: one model for the chance of being retweeted the other for being replied; yet, both on the same feature vectors. All variables except concreteness proofed statistically significant (i.e. with p < 0.001 [13]). Figure 4 displays the feature coefficients for both models. The emotional aspect yielded only limited insight into different triggers for retweets or replies. Yet, it is noteworthy that the coefficients for the group are all negative except arousal. Having a closer look into the parameters for personal orientation we find that the mentioning of other MPs makes a retweet more and a reply less likely. Moreover the use of pronouns is associated with a higher chance of being replied than retweeted. Features in the group link syntax are influencing the two models in exact opposite ways. With the exception of addressing other users (feature accounts) none of the features in that group increases the chance of MPs to receive replies.

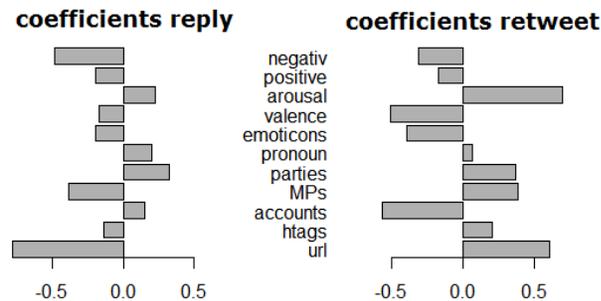

**Figure 3** Regression coefficients of the model predicting reply and retweet chance. (The intercept is not shown here. It is higher for reply because they are more likely in this corpus).

In tackling our guiding question we have seen that the features positively associated with retweets differ from those sparking replies; yet, whether this is just an oddity or representative of an underlying pattern with predictive performance remains to be shown. We test this by measuring the Receiver Operating Characteristic (ROC) [13,14]. The ROC curve tells us about the prediction accuracy by relating the false positive rate with the true positive rate. The bigger the integral or Area Under the Curve the better the model. In our model for predicting the chance of a MPs' tweet receiving a reply the value of the Area Under the Curve is 0.64 and for the retweet model 0.70. So we can assume that the fitted parameters tell us something predictive and are reflecting an underlying pattern.

## Discussion

In our work we investigate if the two modes of responding to tweets (reply and retweet) serve different purposes in the political domain. In this initial study we assume that if both where triggered by a single and not further distinguishable cause we would see a high level of covariance in usage between the two. We tested this by looking at the behavioral changes of the MPs on Twitter themselves over the course of the week as well as on a daily resolution. Next we investigated the textual features of messages by MPs, but now looking at the reaction of the public. In both cases we have seen that the usage of retweets and replies exhibits a distinguishable pattern hinting to that there are indeed separate causes for the two.

Looking at variations over time we found that reply activity seems to be closer associated with typical leisure times such as weekends and after work hours. By containing more link syntax such as external URLs, hashtags and mentioning the public figures of MPs retweets promote one-to-many communication; yet, the retweet-mechanism is shy of reproducing the mentioning of average users, a fact reflected by the accounts coefficients. We assume that with retweeting a wider audience is professionally addressed while replying seems to be a more personal and intimate action.

However, this is just a report on early-stage work, directed at eliciting necessary feedback. Obviously it still covers only a very limited timeframe and as such the dataset is snapshot in time with limited predictive power for future behavior. Yet, more importantly only a small selection handcrafted features has been considered. The inclusion of additional features might improve the performance of the model and shed light on new relations, but this depends on their choice of design and is highly dependent on the experience of the experimenter. One way we are actively investigating is the use of deep belief nets to train semantically meaningful and interpretable features for a regression model.